\documentstyle[12pt]{article}
\begin{document}

\title{ON THE STABILITY OF THE SHEAR--FREE CONDITION}

\author{ L. Herrera$^1$\thanks{e-mail:
laherrera@cantv.net.ve}, A Di Prisco$^1$\thanks{e-mail:
adiprisc@fisica.ciens.ucv.ve} and J. Ospino$^{2}$\thanks {e-mail:
jhozcrae@usal.es} \\
\small{$^1$Escuela de F\'{\i}sica, Facultad de Ciencias,} \\
\small{Universidad Central de Venezuela, Caracas, Venezuela.}\\
\small{$^2$Area de F\'\i
sica Te\'orica. Facultad de Ciencias,Universidad de Salamanca,}\\
\small{Salamanca, Spain.}\\
}

\maketitle

\begin{abstract}
 The evolution  equation for the shear is reobtained for a  spherically symmetric  anisotropic, viscous  dissipative fluid distribution, which allows  us to investigate conditions for the stability of the shear--free condition. The specific case of  geodesic fluids is considered in detail, showing that the shear--free condition, in this particular case, may be  unstable, the departure from the shear--free condition being controlled by the expansion scalar and a single scalar function defined in terms of the anisotropy of the pressure, the shear viscosity  and the Weyl tensor  or, alternatively, in terms of the anisotropy of the pressure, the dissipative variables and the energy density inhomogeneity.
\end{abstract}

\maketitle

\newpage

\section{Introduction}
The relevance of the shear tensor in the evolution of selfgravitating systems and the consequences emerging from its vanishing has been brought out by many authors (see  \cite{SFCF}--\cite{s4} and references therein). Particular attention deserves the possible role of shear in the violation of the cosmic censorship, leading to the appearance of a naked singularity. This has the 
important implication that the resultant spacetime singularity 
of collapse could become visible to faraway observers in the
universe (see \cite{s3p} for a detailed discussion). Furthermore as it has been recently shown \cite{exp} the  shear--free flow (in the nondissipative case)  appears to be equivalent to the well known homologous evolution. It should be recalled that   homology conditions are of great relevance in astrophysics  \cite{astr1}--\cite{astr3}.

Accordingly, it is quite pertinent to ask  under which conditions the shear--free condition is stable. This question in fact entails two different (but related)  questions, namely:
\begin{itemize}
\item
Under which conditions an initially shear--free flow remains shear--free all along the evolution ?
\item Under which conditions a fluid with a small initial shear evolves, keeping always a small value of shear (Lyapunov stability)?
\end{itemize}

Now, in the study of self--gravitating compact objects it is usually assumed that deviations from spherical
symmetry are likely to be incidental rather than  basic features of the process involved. Thus, since  the seminal paper by
Oppenheimer and Snyder \cite{Opp}, most of the work dedicated to the problem of general relativistic gravitational collapse, deal with spherically symmetric fluid distribution.
Accordingly we shall consider spherically symmetric fluid distributions.

The purpose of this work is to provide answers (at least partial) to the above questions. Obviously for doing so we need an evolution equation for the shear. Such an equation was derived by Ellis \cite{ellis1},\cite{ellis2}, for a perfect fluid without any kind of symmetry. Here we shall reobtain that equation for a spherically symmetric  anisotropic viscous dissipative fluid. The physical motivation to consider  such a general fluid has been explained in detail in \cite{1}--\cite{sp} (and references therein). However, since the study of the general case is quite complicated and will probably require the use of numerical techniques, we shall consider in detail the  case of a geodesic fluid.

Nevertheless, even though  the restriction to the geodesic case is dictated by the sake of  simplicity in the analysis, the physical relevance of such a case becomes evident when we recall that Friedman--Lemaitre--Robertson--Walker (FLRW) metrics are shear--free. In other words, it appears that  for a geodesic fluid (without dissipative fluxes),  the stability of the shear--free condition is somehow equivalent to the stability of  FLRW  spacetime itself.

For the geodesic  case it will be shown that departures from the shear--free condition are controlled by the expansion scalar and a scalar function, which is defined in terms of the the Weyl tensor, the anisotropy  of  pressure and the shear viscosity or, alternatively, in terms of pure physical variables. Such a scalar function appears in a natural way in the orthogonal splitting of  the Riemann tensor and is related to the Tolman mass \cite{sp}. Together with the expansion scalar, this scalar function  controls the evolution of the shear.

In the next section we shall present the  general equations and definitions to obtain the evolution equation for the shear. Next in Sec. III that evolution equation is obtained.  The origin and some properties of the  scalar function which plays such an important role in the evolution of the shear are discussed in Sec.IV, and in Sec. V the obtained equation is used to analyze the geodesic case.  Finally,  a discussion of results is presented in the last section.

\section{ENERGY--MOMENTUM TENSOR,  KINEMATICAL VARIABLES AND FIELD EQUATIONS}
We consider a spherically symmetric distribution  of collapsing
fluid, bounded by a spherical surface $\Sigma$. The fluid is
assumed to be locally anisotropic (principal stresses unequal) and undergoing dissipation in the
form of heat flow (to model dissipation in the diffusion approximation), null radiation (to model dissipation in the free streaming approximation) and shearing
viscosity.

Choosing comoving coordinates inside $\Sigma$, the general
interior metric can be written
\begin{equation}
ds^2=-A^2dt^2+B^2dr^2+R^2(d\theta^2+\sin^2\theta d\phi^2),
\label{1}
\end{equation}
where $A$, $B$ and $R$ are functions of $t$ and $r$ and are assumed
positive. We number the coordinates $x^0=t$, $x^1=r$, $x^2=\theta$
and $x^3=\phi$.

The matter energy-momentum $T_{\alpha\beta}$ inside $\Sigma$
has the form
\begin{equation}
T_{\alpha\beta}=(\mu +
P_{\perp})V_{\alpha}V_{\beta}+P_{\perp}g_{\alpha\beta}+(P_r-P_{\perp})\chi_{
\alpha}\chi_{\beta}+q_{\alpha}V_{\beta}+V_{\alpha}q_{\beta}+
\epsilon l_{\alpha}l_{\beta}-2\eta\sigma_{\alpha\beta}, \label{3}
\end{equation}
where $\mu$ is the energy density, $P_r$ the radial pressure,
$P_{\perp}$ the tangential pressure, $q^{\alpha}$ the heat flux,
$\epsilon$ the energy density of the null fluid describing dissipation in the free streaming approximation, $\eta$ the
shear viscosity coefficient, $V^{\alpha}$ the four velocity of the fluid,
$\chi^{\alpha}$ a unit four vector along the radial direction
and $l^{\alpha}$ a radial null four vector. These quantities
satisfy
\begin{equation}
V^{\alpha}V_{\alpha}=-1, \;\; V^{\alpha}q_{\alpha}=0, \;\; \chi^{\alpha}\chi_{\alpha}=1, \;\;
\chi^{\alpha}V_{\alpha}=0, \;\; l^{\alpha}V_{\alpha}=-1, \;\; l^{\alpha}l_{\alpha}=0.
\end{equation}

Observe that we have assumed the shear viscosity  tensor $\pi_{\alpha \beta}$ to satisfy the relation \begin{equation}
\pi_{\alpha \beta}=-2\eta \sigma_{\alpha \beta},
\label{sv}
\end{equation}
where   $\sigma_{\alpha \beta}$ is the shear tensor.  However this last equation is valid only within the context of the standard irreversible thermodynamics (see \cite{8}, \cite{FC} for details).

In a full causal picture of dissipative variables we should not assume (\ref{sv}). Instead, we should  use the  transport equation derived from the corresponding theory (e.g. the M\"{u}ller--Israel--Stewart theory \cite{Muller67}--\cite{131}). However for the sake of simplicity, in this work we shall restrict ourselves to the  standard irreversible thermodynamics  theory.

The acceleration $a_{\alpha}$ and the expansion $\Theta$ of the fluid are
given by
\begin{equation}
a_{\alpha}=V_{\alpha ;\beta}V^{\beta}, \;\;
\Theta={V^{\alpha}}_{;\alpha}. \label{4b}
\end{equation}
and its  shear $\sigma_{\alpha\beta}$ by
\begin{equation}
\sigma_{\alpha\beta}=V_{(\alpha
;\beta)}+a_{(\alpha}V_{\beta)}-\frac{1}{3}\Theta h_{\alpha \beta},\label{4a}
\end{equation}
where $h_{\alpha \beta}=g_{\alpha\beta}+V_{\alpha}V
_{\beta}
.$

We do not explicitly add bulk viscosity to the system because it
can be absorbed into the radial and tangential pressures, $P_r$ and
$P_{\perp}$, of the
collapsing fluid \cite{Chan}.

Since we assumed the metric (\ref{1}) comoving then
\begin{equation}
V^{\alpha}=A^{-1}\delta_0^{\alpha}, \;\;
q^{\alpha}=qB^{-1}\delta^{\alpha}_1, \;\;
l^{\alpha}=A^{-1}\delta^{\alpha}_0+B^{-1}\delta^{\alpha}_1, \;\;
\chi^{\alpha}=B^{-1}\delta^{\alpha}_1, \label{5}
\end{equation}
where $q$ is a function of $t$ and $r$ satisfying $q^\alpha = q \chi^\alpha$.

From  (\ref{4b}) with (\ref{5}) we have for the  acceleration and its scalar $a$,
\begin{equation}
a_1=\frac{A^{\prime}}{A}, \;\; a^2=a^{\alpha}a_{\alpha}=\left(\frac{A^{\prime}}{AB}\right)^2, \label{5c}
\end{equation}
where $a^\alpha= a \chi^\alpha$,
and for the expansion
\begin{equation}
\Theta=\frac{1}{A}\left(\frac{\dot{B}}{B}+2\frac{\dot{R}}{R}\right),
\label{5c1}
\end{equation}
where the  prime stands for $r$
differentiation and the dot stands for differentiation with respect to $t$.
With (\ref{5}) we obtain
for the shear (\ref{4a}) its non zero components
\begin{equation}
\sigma_{11}=\frac{2}{3}B^2\sigma, \;\;
\sigma_{22}=\frac{\sigma_{33}}{\sin^2\theta}=-\frac{1}{3}R^2\sigma,
 \label{5a}
\end{equation}
and its scalar
\begin{equation}
\sigma^{\alpha\beta}\sigma_{\alpha\beta}=\frac{2}{3}\sigma^2,
\label{5b}
\end{equation}
where
\begin{equation}
\sigma=\frac{1}{A}\left(\frac{\dot{B}}{B}-\frac{\dot{R}}{R}\right).\label{5b1}
\end{equation}
Then, the shear tensor can be written as
\begin{equation}
\sigma_{\alpha \beta}= \sigma \left(\chi_\alpha \chi_\beta - \frac{1}{3} h_{\alpha \beta}\right).
\label{sh}
\end{equation}

\subsection{The Einstein equations}

Einstein's field equations for the metric (\ref{1}) are given by
\begin{equation}
G_{\alpha \beta} = 8 \pi T_{\alpha \beta},
\label{Eeq}
\end{equation}
 its non zero components
with (\ref{1}), (\ref{3}) and (\ref{5}) become
\begin{eqnarray}
8\pi T_{00}=8\pi(\mu+\epsilon)A^2
=\left(2\frac{\dot{B}}{B}+\frac{\dot{R}}{R}\right)\frac{\dot{R}}{R}\nonumber\\
-\left(\frac{A}{B}\right)^2\left[2\frac{R^{\prime\prime}}{R}+\left(\frac{R^{\prime}}{R}\right)^2
-2\frac{B^{\prime}}{B}\frac{R^{\prime}}{R}-\left(\frac{B}{R}\right)^2\right],
\label{12} \\
8\pi T_{01}=-8\pi(q+\epsilon)AB
=-2\left(\frac{{\dot R}^{\prime}}{R}
-\frac{\dot B}{B}\frac{R^{\prime}}{R}-\frac{\dot
R}{R}\frac{A^{\prime}}{A}\right),
\label{13} \\
8\pi T_{11}=8\pi
\left(P_r+\epsilon-\frac{4}{3}\eta\sigma\right)B^2 \nonumber\\
=-\left(\frac{B}{A}\right)^2\left[2\frac{\ddot{R}}{R}-\left(2\frac{\dot A}{A}-\frac{\dot{R}}{R}\right)
\frac{\dot R}{R}\right]\nonumber\\
+\left(2\frac{A^{\prime}}{A}+\frac{R^{\prime}}{R}\right)\frac{R^{\prime}}{R}-\left(\frac{B}{R}\right)^2,
\label{14} \\
8\pi T_{22}=\frac{8\pi}{\sin^2\theta}T_{33}=8\pi\left(P_{\perp}+\frac{2}{3}\eta\sigma\right)R^2\nonumber \\
=-\left(\frac{R}{A}\right)^2\left[\frac{\ddot{B}}{B}+\frac{\ddot{R}}{R}
-\frac{\dot{A}}{A}\left(\frac{\dot{B}}{B}+\frac{\dot{R}}{R}\right)
+\frac{\dot{B}}{B}\frac{\dot{R}}{R}\right]\nonumber\\
+\left(\frac{R}{B}\right)^2\left[\frac{A^{\prime\prime}}{A}
+\frac{R^{\prime\prime}}{R}-\frac{A^{\prime}}{A}\frac{B^{\prime}}{B}
+\left(\frac{A^{\prime}}{A}-\frac{B^{\prime}}{B}\right)\frac{R^{\prime}}{R}\right].\label{15}
\end{eqnarray}
\subsection{The mass function}
Let us now introduce the mass function $m(t,r)$ \cite{Misner} (see also \cite{Cahill}), defined by
\begin{equation}
m=\frac{R^3}{2}{R_{23}}^{23}
=\frac{R}{2}\left[\left(\frac{\dot R}{A}\right)^2-\left(\frac{R^{\prime}}{B}\right)^2+1\right].
 \label{17masa}
\end{equation}

Following Misner and Sharp \cite{Misner}, It  is useful to define the proper time derivative $D_T$
given by
\begin{equation}
D_T=\frac{1}{A}\frac{\partial}{\partial t}, \label{16}
\end{equation}
and the proper radial derivative $D_R$,
\begin{equation}
D_R=\frac{1}{R^{\prime}}\frac{\partial}{\partial r}, \label{23a}
\end{equation}
where $R$ defines the areal radius of a spherical surface inside $\Sigma$ ( as
measured from its area).

Using (\ref{16}) we can define the velocity $U$ of the collapsing
fluid  as the variation of the areal radius with respect to proper time, i.e.
\begin{equation}
U=D_TR<0 \;\; \mbox{(in the case of collapse)}. \label{19}
\end{equation}
Then (\ref{17masa}) can be rewritten as
\begin{equation}
E \equiv \frac{R^{\prime}}{B}=\left(1+U^2-\frac{2m}{R}\right)^{1/2}.
\label{20x}
\end{equation}

Using (\ref{12})-(\ref{14}) with (\ref{16}) and (\ref{23a}) we obtain from
(\ref{17masa})
\begin{eqnarray}
D_Tm=-4\pi\left[\left
(\tilde{P}_r-\frac{4}{3}\eta\sigma\right)U+\tilde{q}E\right]R^2,
\label{22Dt}
\end{eqnarray}
and
\begin{eqnarray}
D_Rm=4\pi\left(\tilde{\mu}+\tilde{q}\frac{U}{E}\right)R^2,
\label{27Dr}
\end{eqnarray}
which implies
\begin{equation}
m=4\pi\int^{R}_{0}\left(\tilde{\mu}
+\tilde{q}\frac{U}{E}\right)R^2dR, \label{27intcopy}
\end{equation}
(assuming a regular centre to the distribution, so $m(0)=0$).
Integrating (\ref{27intcopy}) we find
\begin{equation}
\frac{3m}{R^3} = 4\pi\tilde{\mu} - \frac{4\pi}{R^3} \int^R_0{R^3\left(D_R{\tilde \mu}-3 \tilde q \frac{U}{RE}\right) dR}.
\label{3m/R3}
\end{equation}

\section{THE EVOLUTION EQUATION FOR THE SHEAR}
We shall now proceed to deduce the Ellis  evolution equation for the shear, for the specific  fluid distribution discussed in the previous section. For that purpose it will be convenient to express the  energy momentum tensor  (\ref{3})  in the equivalent form
\begin{equation}
T_{\alpha \beta} = \tilde{\mu} V_\alpha V_\beta + \hat{P} h_{\alpha \beta} + \Pi_{\alpha \beta} +
\tilde{q} \left(V_\alpha \chi_\beta + \chi_\alpha V_\beta\right) - 2 \eta \sigma_{\alpha \beta}
\label{Tab}
\end{equation}
with
$$\hat P=\frac{\tilde P_{r}+2P_{\bot}}{3},$$
$$\tilde \mu= \mu+\epsilon,$$
$$\tilde P_{r}=P_r+\epsilon,$$
$$\tilde q= q+\epsilon,$$
$$\Pi=\tilde P_{r}-P_{\bot},$$
$$\Pi_{\alpha \beta}=\Pi\left(\chi_\alpha \chi_\beta - \frac{1}{3} h_{\alpha \beta}\right).$$

Now, Ricci identities for the vector $V_\alpha$ read
\begin{equation}
R^\mu_{\alpha \beta \nu} V_\mu = V_{\alpha; \beta; \nu} - V_{\alpha; \nu; \beta} ,
\label{Ri}
\end{equation}
or, using the well known expression (remember that vorticity vanishes due to the spherical symmetry)
\begin{equation}
V_{\alpha; \mu}=-a_\alpha V_\mu + \sigma_{\alpha \mu} + \frac{1}{3}\Theta h_{\alpha \mu},
\label{deV}
\end{equation}
we obtain
\begin{equation}
\frac{1}{2}R^{\rho}_{\alpha \beta \mu} V_\rho = a_{\alpha;[\beta}V_{\mu]} + a_{\alpha}V_{[\mu;\beta]}+ \sigma_{\alpha[\beta;\mu]} + \frac{1}{3} h_{\alpha[\beta}\Theta_{,\mu]} +\frac{1}{3}\Theta h_{\alpha[\beta;\mu]}.
\label{32}
\end{equation}

Contracting Eq.(\ref{32}) with $V^\beta g^{\alpha \mu}$ , we find the Raychaudhuri equation for the evolution of the expansion
\begin{equation}
\Theta_{;\alpha} V^\alpha + \frac{1}{3}\Theta^2 + \sigma^{\alpha \beta} \sigma_{\alpha \beta} - a^\alpha_{;\alpha} = - V_\rho V^\beta R^\rho_\beta = - 4 \pi (\tilde{\mu} + 3\hat{P}).
\label{Ra}
\end{equation}

On the other hand, contracting (\ref{32}) with $V^\beta h^\alpha_\gamma h^\mu_\nu$ we have
\begin{eqnarray}
&&V^\beta V_\rho R^\rho_{\alpha \beta \mu} h^\alpha_\gamma h^\mu_\nu = h^\alpha_\gamma h^\mu_\nu \left(a_{\alpha;\mu} - V^\beta \sigma_{\alpha \mu;\beta}\right) + a_\gamma a_\nu - \frac{1}{3}V^\beta \Theta_{;\beta} h_{\nu \gamma}\nonumber \\
& - &  h^\mu_\nu V^\beta_{;\mu} \left(\sigma_{\gamma \beta} + \frac{1}{3}\Theta h_{\gamma \beta}\right),
\label{35}
\end{eqnarray}
which by using (\ref{deV}),  can be written as
\begin{eqnarray}
&&V^\beta V_\rho R^\rho_{\alpha \beta \mu} h^\alpha_\gamma h^\mu_\nu = h^\alpha_\gamma h^\mu_\nu \left(a_{\alpha;\mu} - V^\beta \sigma_{\alpha \mu;\beta}\right) + a_\gamma a_\nu  - \frac{1}{3}V^\beta \Theta_{;\beta} h_{\nu \gamma}\nonumber \\
 &-& \frac{2}{3} \Theta \sigma_{\gamma \nu}
- \frac{\sigma^2}{3} \left(\chi_\gamma \chi_\nu + \frac{1}{3}h_{\gamma \nu}\right) - \frac{1}{9} \Theta^2 h_{\gamma \nu}.
\label{35p}
\end{eqnarray}

Next, the
Riemann tensor may be expressed through the Weyl tensor
$C^{\rho}_{\alpha
\beta
\mu}$, the  Ricci tensor $R_{\alpha\beta}$ and the scalar curvature $\cal R$,
as:
$$
R^{\rho}_{\alpha \beta \mu}=C^\rho_{\alpha \beta \mu}+ \frac{1}{2}
R^\rho_{\beta}g_{\alpha \mu}-\frac{1}{2}R_{\alpha \beta}\delta
^\rho_{\mu}+\frac{1}{2}R_{\alpha \mu}\delta^\rho_\beta$$
\begin{equation}
-\frac{1}{2}R^\rho_\mu g_{\alpha
\beta}-\frac{1}{6}{\cal R}(\delta^\rho_\beta g_{\alpha \mu}-g_{\alpha
\beta}\delta^\rho_\mu).
\label{34}
\end{equation}
Contracting  (\ref{34}) with$ V_{\rho} V^\beta
h^\alpha_\gamma h^\mu_\nu$ and using  Einstein equations, we find:
\begin{equation}
V_{\rho} V^\beta h^\alpha_\gamma h^\mu_\nu R^{\rho}_{\alpha \beta \mu} = E_{\gamma \nu } + \frac{4 \pi}{3} \left(\tilde{\mu} + 3\hat{P}\right)h_{\gamma \nu} - 4 \pi \Pi_{\gamma \nu} + 8 \pi \eta \sigma_{\gamma \nu}, \nonumber \\
\label{37}
\end{equation}
where $E_{\gamma \nu }$ denotes the ``electric'' part of the Weyl tensor (in the spherically symmetric case the ``magnetic'' part of the Weyl tensor 
vanishes), and is defined by
\begin{equation}
E_{\alpha \beta} = C_{\alpha \mu \beta \nu} V^\mu V^\nu,
\label{elec}
\end{equation}
whose non trivial components are
\begin{eqnarray}
E_{11}&=&\frac{2}{3}B^2 {\cal E},\nonumber \\
E_{22}&=&-\frac{1}{3} R^2 {\cal E}, \nonumber \\
E_{33}&=& E_{22} \sin^2{\theta},
\label{ecomp}
\end{eqnarray}
where
\begin{eqnarray}
{\cal E}&= &\frac{1}{2 A^2}\left[\frac{\ddot R}{R} - \frac{\ddot B}{B} - \left(\frac{\dot R}{R} - \frac{\dot B}{B}\right)\left(\frac{\dot A}{A} + \frac{\dot R}{R}\right)\right]\nonumber \\
&+& \frac{1}{2 B^2} \left[\frac{A^{\prime\prime}}{A} - \frac{R^{\prime\prime}}{R} + \left(\frac{B^{\prime}}{B} + \frac{R^{\prime}}{R}\right)\left(\frac{R^{\prime}}{R}-\frac{A^{\prime}}{A}\right)\right] - \frac{1}{2 R^2}.
\label{E}
\end{eqnarray}

 Observe that  the electric part of
Weyl tensor, may be written as:
\begin{equation}
E_{\alpha \beta}={\cal E} (\chi_\alpha \chi_\beta-\frac{1}{3}h_{\alpha \beta}).
\label{52}
\end{equation}

Using  (\ref{Ra}), (\ref{37}) can be written as
\begin{equation}
V_{\rho} V^\beta h^\alpha_\gamma h^\mu_\nu R^{\rho}_{\alpha \beta \mu} = E_{\gamma \nu }
- \left(\frac{1}{3} V^{\alpha} \Theta_{;\alpha} + \frac{\Theta^2}{9} + \frac{2}{9}\sigma^2 - \frac{a^\alpha_{;\alpha}}{3}\right)h_{\gamma  \nu}
- 4 \pi \Pi_{\gamma \nu} + 8 \pi \eta \sigma_{\gamma \nu},
\label{37p}
\end{equation}
then, from Eq.(\ref{35p}) and Eq.(\ref{37p}) it follows
\begin{eqnarray}
&&E_{\gamma \nu } - 4 \pi \Pi_{\gamma \nu} + 8 \pi \eta \sigma_{\gamma \nu}  = \nonumber \\
&& h^\alpha_\gamma h^\mu_\nu \left(a_{\alpha;\mu} - V^\beta \sigma_{\alpha \mu;\beta}\right) -
 \frac{a^\alpha_{;\alpha}}{3}h_{\gamma \nu} + a_\gamma a_\nu - \frac{1}{3}\sigma_{\gamma \nu} \left(2 \Theta + \sigma\right)
\label{38}
\end{eqnarray}
or, using (\ref{sh})
\begin{eqnarray}
&&E_{\gamma \nu } - 4 \pi \Pi_{\gamma \nu} + 8 \pi \eta \sigma_{\gamma \nu}  = \nonumber \\
&& h^\alpha_\gamma h^\mu_\nu \left(a_{\alpha;\mu} - V^\beta \sigma_{\alpha \mu;\beta}\right)
-  \frac{a^\alpha_{;\alpha}}{3}h_{\gamma \nu}+ a_\gamma a_\nu   - \frac{2}{3}\sigma_{\gamma \nu} \Theta
+\frac{2}{9} \sigma^2 h_{\gamma \nu} - \sigma^\beta_\nu \sigma _{\gamma \beta}.\nonumber \\
\label{38p}
\end{eqnarray}

Finally, contracting (\ref{38p}) with $\chi^\gamma \chi^\nu$ we obtain
\begin{equation}
{\cal E} - 4\pi \Pi + 8 \pi \eta \sigma = a^\dagger - \sigma^\ast +a^2 - \frac{\sigma^2}{3} - \frac{2}{3} \Theta \sigma - a \frac{R'}{RB}\;,
\label{shev}
\end{equation}
with $f^\dagger=f_{,\alpha}\chi^\alpha$ and
$f^\ast=f_{,\alpha}V^\alpha$.

Using (\ref{5}), (\ref{shev}) takes the form
\begin{equation}
Y_{TF}\equiv {\cal E} - 4\pi \Pi + 8 \pi \eta \sigma = \frac{a^\prime}{B} - \frac{\dot{\sigma}}{A} +a^2 - \frac{\sigma^2}{3} - \frac{2}{3} \Theta \sigma - a \frac{R'}{RB}\;.
\label{shevp}
\end{equation}
This is the evolution equation for the shear we were looking for (we recall that there is only one independent component of the shear tensor  in our case). It is  equivalent (in comoving coordinates) to Eq.(71) in \cite{9}  or Eq.(101) in \cite{sp} (notice that $E$ in those references equals $-{\cal E}$ here). 

We shall    use the above equation to study the stability of the shear--free condition for the case of a geodesic fluid. However before doing that we shall discuss about the origin and the physical properties of  $Y_{TF}$.
\section{ON THE ORIGIN AND PROPERTIES OF $Y_{TF}$}
As we mention in the Introduction, the scalar function $Y_{TF}$ appears in a natural way in the orthogonal splitting of the Riemann tensor (see \cite{sp} for details).

Indeed, the electric part of the Riemann tensor (which is one of the element of that splitting) is defined by \cite{bel1}, \cite{parrado}
\begin{equation}
Y_{\alpha \beta}=R_{\alpha \gamma \beta \delta}V^{\gamma}V^{\delta},
\label{electric}
\end{equation}
which in turn  can be splitted in terms of its trace and  the corresponding trace--free tensor, i.e.
\begin{equation}
Y_{\alpha \beta}=\frac{1}{3}Tr Y h_{\alpha \beta}+ Y_{<\alpha \beta>},
\label{esn}
\end{equation}
with $Tr Y=Y^\alpha_\alpha$ and,
\begin{equation}
Y_{<\alpha \beta>}=h^\mu_\alpha h^\nu_\beta\left(Y_{\mu \nu}-\frac{Tr Y}{3} h_{\mu \nu}\right),
\label{esnII}
\end{equation}
this last tensor may also be written as
\begin{equation}
Y_{<\alpha  \beta>}=Y_{TF}\left(\chi_\alpha \chi_ \beta-\frac{1}{3} h_{\alpha \beta}\right).
\label{esnIII}
\end{equation}

To obtain $Tr Y$ and $Y_{TF}$ we may proceed as in \cite{sp} or, directly from (\ref{electric})--(\ref{esnIII}),  using  the expressions for  the Riemann tensor components in terms of the Einstein tensor components,  given in the Appendix of \cite{13}.  Either way  the result is
\begin{equation}
TrY\equiv Y_T=4\pi\left(\tilde \mu+3\tilde P_r-2\Pi \right),
\label{esnV}
\end{equation}
and
  \begin{equation}
Y_{TF}=4 \pi(-\Pi+2\eta \sigma)+\cal{E}.
\label{defYTF}
\end{equation}
Next, using  (\ref{12}), (\ref{14}), (\ref{15}) with (\ref{17masa}) and (\ref{E}) we obtain
\begin{equation}
\frac{3m}{R^3}=4\pi \left(\tilde{\mu}-\Pi+2\eta\sigma\right) - \cal{E},
\label{mE}
\end{equation}
which combined with (\ref{3m/R3})  and (\ref{defYTF}) produces

\begin{equation}
Y_{TF}= -8\pi\Pi + 16\pi\eta\sigma+\frac{4\pi}{R^3}\int^R_0{R^3\left(D_R \tilde{\mu}-3\tilde{q}\frac{U}{RE}\right)dR}.
\label{Y}
\end{equation}

Thus the scalar $Y_{TF}$ may be expressed through the Weyl tensor and the anisotropy of pressure  or in terms of the anisotropy of pressure, the density inhomogeneity and  the dissipative variables. It is worth recalling that a link between $Y_{TF}$ and the Tolman mass has been established in \cite{sp}.

We shall now bring out  the role of this scalar function in the stability of the shear--free condition for the geodesic fluid.

\section{THE GEODESIC FLUID}
If we assume the fluid to be geodesic (i.e. $a^\mu=0=a$) then it follows from (\ref{5c}) and rescaling $t$, that   $A=1$. In this case (\ref{shevp}) reads
\begin{equation}
\dot \sigma+\frac{\sigma^2}{3}+\frac{2\Theta \sigma}{3}+Y_{TF}=0,
\label{111}
\end{equation}
where $Y_{TF}$ is defined through (\ref{defYTF}) or (\ref{Y}).

Let us first assume that $Y_{TF}=0$, then the general solution of (\ref{111}) is

\begin{equation}
\sigma(r,t) =\frac{3c(r)}{\left[c(r) t -e^{\frac{2}{3}\int\Theta dt}\left(1-\frac{2}{3}c(r)\int \Theta te^{-\frac{2}{3}\int\Theta d\tilde t}dt\right)\right]}\; ,  \label{222}
\end{equation}
 where $c(r)$ is a function of integration. After some simple manipulations, this solution can be rewritten as
 \begin{equation}
 \sigma(r,t)=\frac{3 c(r) e^{-(2/3)\int{\Theta dt}}}{c(r) \int{e^{-(2/3)\int{\Theta d\tilde t}}dt}-1}.
 \label{solJ}
 \end{equation}

 If we now demand the shear to vanish initially, (i.e. $\sigma(0,r)=0$) then as it follows from (\ref{222}) or (\ref{solJ}) we must have (assuming $\Theta$ is a regular function of its arguments)

\begin{equation}
c(r)=0, \label{333}
\end{equation}
implying $\sigma=0$ for all $t$.
Therefore, if $Y_{TF}=0$, {\it  the only solution compatible with an initially shear--free flow is a shear--free flow}. This result is also evident from (\ref{111}).

 Before proceeding further, the following remark is in order: Observe that the condition $Y_{TF}=0$ does not imply conformal flatness (${\cal E}=0$) as it is obvious from (\ref{defYTF}), unless we assume further the fluid to be perfect. In this latter case the result above is compatible with the one  obtained by Stephani \cite{s1}, \cite{s2} stating that the most general conformally  flat perfect fluid solution differs from FLRW only by having nonzero acceleration.

Let us now consider the case (always with $Y_{TF}=0$) when,  initially, the flow is close to the shear--free condition, but still $\sigma \neq0$. Then the question is under which conditions the fluid will evolve, keeping close to the shear--free regime (Lyapunov stability).

Thus, let us assume
\begin{equation}
\sigma(0,r)=\epsilon \bar \sigma(r), \qquad |\epsilon|<<1
\label{444}
\end{equation}

Now, for sufficiently small $\epsilon$ and assuming $\Theta$ to be a regular function of $t$, we obtain from (\ref{222})
\begin{equation}
c(r) \approx  {\cal O}(\epsilon), \label{5555p}
\end{equation}
implying

\begin{equation}
\sigma \approx -\epsilon e^{-\frac{2}{3}\int \Theta dt}.\label{5555}
\end{equation}

From the above it is evident  that if $\Theta>0$ then the shear will remain always close to the initial (``quasi--shear--free'' condition). However it is also evident that for $\Theta<0$ the situation radically changes and in principle departures from the quasi--shear--free condition may be expected, depending on $\Theta$.

Let us now consider the case  $Y_{TF}\neq 0$.

We shall first  assume that $Y_{TF}$ (at least initially)  is small, i.e.
\begin{equation}
Y_{TF}=\alpha \bar{Y}_{TF}; \qquad \alpha <<1,
\label{alX}
\end{equation}
and
\begin{equation}
\sigma(t,r) = \sigma_o(t,r) + \beta \sigma_1(t,r), \qquad \beta <<1
\label{sig}
\end{equation}
\begin{equation}
\Theta(t,r) = \Theta_o(t,r) + \xi \Theta_1(t,r), \qquad \xi <<1
\label{the}
\end{equation}

where $\sigma_o(t,r)$ and  $\Theta_o(t,r)$ correspond to the general solution of (\ref{111}) for $Y_{TF}=0$ given by  (\ref{222}).

Introducing (\ref{sig}) and (\ref{the}) into (\ref{111}) and linearizing we obtain for $\sigma_1(t,r)$ the following equation
\begin{equation}
\dot{\sigma}_1 + \frac{2}{3}\left(\sigma_o + \Theta_o\right) \sigma_1 + \frac{2\xi}{3\beta} \Theta_1 \sigma_o + \frac{\alpha}{\beta} \bar{Y} _{TF}= 0,
\label{eqsigx}
\end{equation}
which may be easily  integrated to obtain
\begin{eqnarray}
\sigma_1(t,r)&=& f(r) e^{-(2/3)\int{(\sigma_o+\Theta_o) dt}} \nonumber\\
&-&e^{-(2/3)\int{(\sigma_o+\Theta_o) dt}} \int{\frac{1}{\beta}\left(\frac{2}{3}\xi \Theta_1 \sigma_o + \alpha \bar{Y} _{TF}\right) e^{(2/3)\int{(\sigma_o+\Theta_o) d\tilde{t}}}  dt},\nonumber\\
\label{sigx}
\end{eqnarray}
where $f(r)$ is an integration function.
Introducing this last equation into (\ref{sig}) we have
\begin{eqnarray}
\sigma(t,r)&=& \sigma_o(t,r) + \beta f(r) e^{-(2/3)\int{(\sigma_o+\Theta_o) dt}} \nonumber\\
&-&e^{-(2/3)\int{(\sigma_o+\Theta_o) dt}} \int{\left(\frac{2}{3}\xi \Theta_1 \sigma_o +\alpha \bar{Y} _{TF}\right) e^{(2/3)\int{(\sigma_o+\Theta_o) d\tilde{t}}}  dt}.\nonumber\\
\label{sigf}
\end{eqnarray}

Let us now assume that the fluid is initially shear--free, which implies $\sigma_0=0$. Then it follows at once from (\ref{sigf})
\begin{eqnarray}
\sigma(t,r)&=& \beta f(r) e^{-(2/3)\int{\Theta_o  dt}} \nonumber\\
&-&e^{-(2/3)\int{\Theta_o dt}} \int{\alpha \bar{Y} _{TF} e^{(2/3)\int{\Theta_o d\tilde{t}}}  dt}.\nonumber\\
\label{sigfIII}
\end{eqnarray}
If, $Y_{TF}=0$ ($\alpha=0$) then from the condition of an initially shear--free fluid, we have that $f(r)=0$, implying $\sigma(t,r)=0$ for all $t$, which is the result obtained before.

However, for any $\alpha \neq 0$, it appears from (\ref{sigf}) that {\it the fluid may evolve to a non--vanishing shear regime, for any sign of $\Theta$, depending on the specific time dependence of $\bar Y_{TF}$}. In other words, even  for small values of $\alpha$ the fluid may deviate from its initial shear--free condition.

If the fluid initially satisfies the quasi--shear--free condition, which implies that $\sigma_0$ is small but non--vanishing. Then in the case $Y_{TF}=\alpha=0$ we obtain as before that for $\Theta<0$  departures from the quasi--shear--free condition may be expected, depending on $\Theta$.

In the general ($\alpha \neq 0$) case  however, departures from the quasi--shear--free condition may be expected  along the evolution, depending on $Y_{TF}$ for any sign of $\Theta$.

An alternative (and useful) expression for the shear, which allows to study the same problem, may be obtained as follows.

\noindent From the geodesic condition
\begin{equation}
a=0\Rightarrow A=1, \label{aqed}
\end{equation}
the field equation (\ref{13}) becomes
\begin{equation}
\frac{\dot R ^\prime}{R}-\frac{\dot B}{B}\frac{R^\prime}{R}=4\pi \tilde{q} B,
\label{qed}
\end{equation}
or,
\begin{equation}
\frac{\dot B}{B}= \frac{\dot R^\prime}{R^\prime} - \frac{F R}{R^\prime},
\label{qedb}
\end{equation}
where $F\equiv 4\pi\tilde{q}B$.

Then,   (\ref{qedb}) with  (\ref{5c1}) and (\ref{5b1}) produces
\begin{equation}
2\Theta+\sigma=3\left(\frac{\dot R^{\prime}}{R^\prime}+\frac{\dot
R}{R}-\frac{F R}{R^\prime}\right)=3(ln(R^ \prime R)\dot) - 3 \frac{F R}{R^\prime}
\label{thsd}
\end{equation}
\noindent Introducing (\ref{thsd}) into (\ref{111}) we get
\begin{equation}
\dot \sigma+\left[(ln(R^\prime R)\dot )-\frac{F R}{R^\prime}\right]\sigma +Y_{TF}=0,
\label{seqd}
\end{equation}
\noindent which after integration yields
\begin{equation}
\sigma=\frac{e^{\int{(FR/R^\prime)dt}}}{RR^\prime}\left(-\int R^\prime R Y_{TF}e^{-\int{(FR/R^\prime)d\tilde{t}}}dt+C(r)\right).
\label{seqd1}
\end{equation}

In the non--dissipative case ($F=0$), (\ref{seqd1}) takes the very simple form

\begin{equation}
\sigma=\frac{1}{RR^\prime}\left(-\int R^\prime R Y_{TF}dt+C(r)\right).
\label{seq1}
\end{equation}

Now, if we assume that $Y_{TF}=0$ then it follows at once from (\ref{seqd1})  that for  an initially shear--free fluid we must have $C(r)=0$ implying $\sigma(t,r)=0$.

Also, for an initially quasi--shear--free fluid (with   $Y_{TF}=0$) it should be clear that departures from that initial condition are possible if  $\Theta<0$ and such that it produces  a sharp decreasing of $R$ (remember that in order to avoid shell crossing singularities we must have $R^\prime>0$). Thus we recover  our previous result.

In the general case ($Y_{TF}\neq 0$) it is obvious from (\ref{seqd1}) that departures from either the shear--free or the quasi--shear--free initial conditions are possible.

\section{DISCUSSION}
We have carried out a study on the stability of the shear--free condition based on  the evolution equation for the shear  presented  in Sec. III . As shown in the previous section, a major role in such study is played  by the scalar $Y_{TF}$. This scalar, which   appears in the orthogonal splitting of the Riemann tensor, is not only related with the Tolman mass as shown in \cite{sp}, but may be expressed through purely physical quantities as in (\ref{Y}).

We shall not insist on the relevance of the question considered in this work (the stability of the shear--free condition) since that was clearly stated in the Introduction. However we would like to comment further on the results concerning the geodesic case.

Indeed, as mentioned before, it is worth noticing that  in the non--dissipative, isotropic case, it may be easily shown that (e.g. see \cite{9})
\begin{equation}
Y_{TF}\equiv {\cal{E}}= 0 \Leftrightarrow \sigma =0.
\label{impl}
\end{equation}

Thus for that particular case (geodesic, non--dissipative, isotropic pressure) the shear makes the difference between FLRW metrics ($Y_{TF}=\sigma={\cal{E}}=0$) and LTB  metrics ($Y_{TF}\neq 0, \sigma \neq 0$). 

Therefore,  the discussion in Sec.V illustrates for this particular case, how departures from an initial   FLRW  spacetime   are controlled by $Y_{TF}$.
 
Finally, it should be emphasized  the fact that, even though  it may be intuitively obvious that dissipative processes, local anisotropy of pressure and  energy density inhomogeneity should affect the stability of the shear--free condition, it is not so obvious  that  the above mentioned factors affect  the stability of the shear--free condition  only through  their specific combination  given by (\ref{Y}). Thus for example we could  consider the peculiar case when all those factors are present but  they cancel each other in (\ref{Y}), producing $Y_{TF}=0$, in such a case the shear--free condition  would be stable in spite of the fact that the fluid is non--homogeneous, non--isotropic and non--dissipative.
 
\section*{Acknowledgments.}
LH and
ADP wish to thank Universite Paris VI, France, Universitat de Les Illes Balears, Spain and Universidad de Salamanca,
Spain for their hospitality. ADP also wishes to thank Comisi\'on de Investigaci\'on and Comisi\'on de Estudios de Posgrado,
 Facultad de Ciencias, Universidad Central de Venezuela, Venezuela, for finantial support. LH also  wishes to thank Fundacion Empresas
 Polar for finantial support.
JO acknowledges financial support from the Universidad  de  Salamanca
(Spain)
 under grant USAL2008A11.

\end{document}